\documentclass[a4,aps,prd,groupedaddress,showpacs,nofootinbib]{revtex4}
\usepackage{graphicx,amssymb,amsmath}
\usepackage[english]{babel}

\newcommand{\be}{\begin{equation}}\newcommand{\ee}{\end{equation}}
\newcommand{\bea}{\begin{eqnarray}}\newcommand{\eea}{\end{eqnarray}}
\newcommand{\beaa}{\begin{eqnarray}}\newcommand{\eeaa}{\end{eqnarray}}
\newcommand{\ba}{\begin{array}}\newcommand{\ea}{\end{array}}
\newcommand{\bit}{\begin{itemize}}\newcommand{\eit}{\end{itemize}}
\newcommand{\ben}{\begin{enumerate}}\newcommand{\een}{\end{enumerate}}

\def\lan{\langle}
\def\lf{\left}

\def\non{\nonumber}
\def\ran{\rangle}
\def\rar{\rightarrow}
\def\ri{\right}

\def\al{\alpha}\def\bt{\beta}

\def\te{\theta}

\def\1{{_{1}}}\def\2{{_{2}}}
\def\nof{:\;\!\!\;\!\!:}

\begin{document}

\title{A new perspective in the dark energy puzzle from particle mixing phenomenon}

\author{M.Blasone${}^{\flat}$, A.Capolupo${}^{\flat}$, S.Capozziello${}^{\sharp}$, G.Vitiello${}^{\flat}$}


\affiliation{ ${}^{\flat}$
Dipartimento di Matematica e Informatica,
 Universit\`a di Salerno and Istituto Nazionale di Fisica Nucleare,
 Gruppo Collegato di Salerno, 84100 Salerno, Italy,
\\ ${}^{\sharp}$ Dipartimento di Scienze Fisiche, Universit\`a di Napoli "Federico II" and INFN Sez. di Napoli,
Compl. Univ. Monte S. Angelo, Ed.N, Via Cinthia, I-80126 Napoli,
Italy.}


\vspace{2mm}

\begin{abstract}

We report on recent results on particle mixing and oscillations in
quantum field theory. We discuss the role played in cosmology by the
vacuum condensate induced by the neutrino mixing phenomenon. We show
that it can contribute to the dark energy of the universe.

\end{abstract}


\maketitle

\section{Introduction}

The study of neutrino mixing in the context of quantum field theory
(QFT)
\cite{BV95,Fujii:1999xa,JM01,hannabuss,yBCV02,Capolupo:2004av} and
the progresses in the understanding of such a phenomenon
\cite{Blasone:2005ae,Blasone:2006jx}, together  with
 the definitive experimental proof of neutrino oscillations
\cite{SNO,K2K}, open new scenarios to the research in fundamental
physics. Indeed, it has emerged an unexpectedly rich
non-perturbative structure associated to the mixing of neutrino (as
well as boson \cite{BCRV01}) fields, hidden in the vacuum for the
flavor fields. This has been shown to be a condensate of massive
neutrino-antineutrino pairs. Several consequences of this discovery
have been analyzed, including the effects on flavor oscillation
formulas \cite{BV95,yBCV02} and the implications in cosmology and
astro-particle physics
\cite{Blasone:2004yh,Capolupo:2006et,Mavromatos:2007sp}.

In this review we show that the energy content of the neutrino
mixing vacuum condensate \cite{Blasone:2004yh} can represent a
component of dark energy \cite{Capolupo:2006et} that, at present
epoch, assumes the behavior and the value of the observed
cosmological constant. We compute such a value and show that, above
a threshold, it is slowly diverging and its derivative with respect
to the cut-off value goes actually to zero, which allows to use the
cut-off at its electroweak scale value, provided one limits himself
to considering the two (lighter) generations in neutrino mixing. In
the case one includes mixing with heavier neutrinos, a value of the
dark energy compatible with its upper bound is obtained for a
cut-off of the order of the natural scale of neutrino mixing. As  we
will show in a forthcoming paper \cite{Capolupo2008},  the use of
such an infrared momentum cut-off is motivated, at the present
epoch, by the negligible breaking of Lorentz invariance due to the
vacuum condensate caused by neutrino mixing .

The remarkable improvement of the dark energy
value computed in the present paper, with respect to the disagreement of $123$
orders of magnitude in standard approaches \cite{Sahni:2004ai},
makes the present treatment worth to be discussed.

Our result links together dark energy with the sub-eV neutrino mass
scale \cite{Maltoni:2004ei}. The link comes from the
neutrino-antineutrino pair vacuum condensate.

We point out that our work differs from the approach
\cite{deVega:2007rh} based on vacuum contributions from light
particles like neutrinos and axions. In the present review we do not
resort to axion contributions. Nevertheless, it is in our future
plan to compare the two approaches in order to clarify the
differences and the similarities. Moreover, the non-perturbative
feature here presented leads us to believe that a
neutrino--antineutrino asymmetry, if any, related with lepton number
violation \cite{Boyanovsky:2004xz} would not affect much our
result. Also this point deserves to be better clarified, which we
plan to do in a subsequent work. Finally, we observe that the
non-perturbative contribution discussed in the present review is of
different origin with respect to the vacuum energy contribution of
massive spinor fields arising from a radiative correction at some
perturbative order \cite{Coleman:1973}.

The review is organized as follows. In Section II, we introduce
the neutrino mixing formalism in QFT. In Section
III we present the neutrino mixing contribution to the dark energy
of the universe. Conclusions are drawn in
Section IV.

\section{Neutrino mixing in Quantum Field Theory}

The neutrino mixing phenomenon was firstly studied in the context of
quantum mechanics  (QM)
\cite{Pontecorvo:1957cp,Maki:1962mu,Fujii64,Gribov:1968kq,Bilenky:1978nj,Bilenky,Mohapatra:1991ng,Wolfenstein:1977ue}
and subsequently analyzed in the framework of the QFT formalism
\cite{BV95,Fujii:1999xa,JM01,hannabuss,yBCV02,Capolupo:2004av,Blasone:2005ae,Blasone:2006jx},
which we shortly summarize in the following (for a detailed review
see \cite{Capolupo:2004av}).

In the very effective Pontecorvo's formalism neutrino mixing is
considered from the standpoint of QM and the attention is focused on
the mixing of ''states''. The fact that neutrinos are actually
described by field operators is completely neglected. The reason for
that is the necessity of the effectiveness of the formalism which is
required to readily fit the experimental search for mixing and
oscillations. Therefore, any simplification of the matter to be
treated is adopted, provided the resulting description would be
sufficiently accurate and descriptive/predictive of the experimental
observations. As a matter of fact, the successive development of the
experimental search has been supporting such an attitude. From a
theoretical point of view, there is, however, the necessity to
understand how mixing and oscillations can be properly described in
the realm of QFT, which provides anyway the proper setting for
neutrino dynamics, as known since the birth of QFT. It is then also
necessary to understand how the correct formalism connects to the
Pontecorvo's approximate scheme. This has been indeed the program of
the research line which has led to the QFT formulation of the
neutrino mixing and oscillation. Such a program has been
successively extended so to incorporate other particle mixing (quark
mixing, boson mixing). Here we only sketch the skeleton of the QFT
mixing formalism and to do that we consider two neutrinos. Extension
to three neutrino (to any number of generations, in principle) is in
the literature \cite{yBCV02}. The reader who wants the guaranties
offered by a rigorous mathematical proof of our treatment may
usefully read the papers in Refs.\cite{hannabuss}.

The mixing transformations for two Dirac neutrino fields are
\begin{eqnarray} \nonumber\label{mix}
\nu_{e}(x) &=&\nu_{1}(x)\,\cos\theta + \nu_{2}(x)\,\sin\theta
\\
\nu_{\mu}(x) &=&-\nu_{1}(x)\,\sin\theta + \nu_{2}(x)\,\cos\theta
\;,\end{eqnarray}
where $\nu_{e}(x)$ and $\nu_{\mu}(x)$ are the
fields with definite flavors, $\theta$ is the mixing angle and
$\nu_1$ and $\nu_2$ are the fields with definite masses $m_{1} \neq m_{2}$:
\bea\label{freefi}
 \nu _{i}(x)=\frac{1}{\sqrt{V}}{\sum_{{\bf k} ,
r}} \left[ u^{r}_{{\bf k},i}\, \al^{r}_{{\bf k},i}(t) +
v^{r}_{-{\bf k},i}\, \bt^{r\dag}_{-{\bf k},i}(t) \ri] e^{i {\bf
k}\cdot{\bf x}},\qquad \, \qquad i=1,2,
\eea
with  $ \al_{{\bf k},i}^{r}(t)=\al_{{\bf
k},i}^{r}\, e^{-i\omega _{k,i}t}$,
$ \bt_{{\bf k},i}^{r\dag}(t) = \bt_{{\bf k},i}^{r\dag}\, e^{i\omega_{k,i}t},$
and $ \omega _{k,i}=\sqrt{{\bf k}^{2} + m_{i}^{2}}.$
The operators $\alpha ^{r}_{{\bf k},i}$ and $ \beta ^{r }_{{\bf k},i}$, $
i=1,2 \;, \;r=1,2$ annihilate the vacuum
state $|0\rangle_{1,2}\equiv|0\rangle_{1}\otimes |0\rangle_{2}$:
$\alpha ^{r}_{{\bf k},i}|0\rangle_{12}= \beta ^{r }_{{\bf
k},i}|0\rangle_{12}=0$.
 The anticommutation relations are:
$\left\{ \nu _{i}^{\alpha }(x),\nu _{j}^{\beta
\dagger }(y)\right\} _{t=t^{\prime }}=\delta ^{3}({\bf x-y})\delta
_{\alpha \beta } \delta _{ij},$ with $\alpha ,\beta =1,...4,$ and $\left\{ \alpha _{{\bf k},i}^{r},\alpha _{{\bf
q},j}^{s\dagger }\right\} =\delta _{{\bf kq}}\delta _{rs}\delta
_{ij};$ $\left\{ \beta _{{\bf k},i}^{r},\beta _{{\bf
q,}j}^{s\dagger }\right\} =\delta _{{\bf kq}}\delta _{rs}\delta
_{ij},$ with $i,j=1,2.$
All other anticommutators are zero. The orthonormality and
completeness relations are:
$u_{{\bf k},i}^{r\dagger }u_{{\bf k},i}^{s} = v_{{\bf
k},i}^{r\dagger }v_{{\bf k},i}^{s} = \delta _{rs},\; $
$u_{{\bf k},i}^{r\dagger }v_{-{\bf k},i}^{s} = v_{-{\bf k}
,i}^{r\dagger }u_{{\bf k},i}^{s} = 0,\;$
and
$\sum_{r}(u_{{\bf k},i}^{r}u_{{\bf k},i}^{r\dagger
}+v_{-{\bf k},i}^{r}v_{-{\bf k},i}^{r\dagger }) = 1.$

The mixing transformation
Eqs.(\ref{mix}) can be written as \cite{BV95}:
\bea \label{mixG} \nu_{e}^{\alpha}(x) = G^{-1}_{\bf \te}(t)\;
\nu_{1}^{\alpha}(x)\; G_{\bf \te}(t) \\ \non \nu_{\mu}^{\alpha}(x)
= G^{-1}_{\bf \te}(t)\; \nu_{2}^{\alpha}(x)\; G_{\bf \te}(t) \eea
where the mixing generator $G_{\bf \te}(t)$ is given by
\bea\label{generator12} G_{\bf \te}(t) = exp\left[\theta \int
d^{3}{\bf x} \left(\nu_{1}^{\dag}(x) \nu_{2}(x) - \nu_{2}^{\dag}(x)
\nu_{1}(x) \right)\right]\,.
\eea
At finite volume, $G_{\bf \te}(t)$ is an unitary operator,
$G^{-1}_{\bf \te}(t)=G_{\bf -\te}(t)=G^{\dag}_{\bf \te}(t)$,
preserving the canonical anticommutation relations; $G^{-1}_{\bf
\te}(t)$ maps the Hilbert spaces for $\nu_{1}$ and $\nu_{2}$ fields ${\cal H}_{1,2}$ to
the Hilbert spaces for flavored fields ${\cal H}_{e,\mu}$: $
G^{-1}_{\bf \te}(t): {\cal H}_{1,2} \mapsto {\cal H}_{e,\mu}.$ In
particular, for the vacuum $|0 \rangle_{1,2}$ we have, at finite
volume $V$:
\bea\label{flavvac}
 |0(t) \rangle_{e,\mu} = G^{-1}_{\bf \te}(t)\;
|0 \rangle_{1,2}\,.
\eea
$|0 \rangle_{e,\mu}$ is the vacuum for ${\cal H}_{e,\mu}$, which
we will refer to as the flavor vacuum.
The explicit expression for $|0\rangle_{e,\mu}$  at time $t=0$ in
the reference frame for which ${\bf k}=(0,0,|{\bf k}|)$ is
\bea
\non\label{0emu} |0\rangle_{e,\mu} &=& \prod_{r,{\bf k}}
\Big[(1-\sin^{2}\theta\;|V_{{\bf k}}|^{2})
-\epsilon^{r}\sin\theta\;\cos\theta\; |V_{{\bf k}}|
(\alpha^{r\dag}_{{\bf k},1}\beta^{r\dag}_{-{\bf k},2}+
\alpha^{r\dag}_{{\bf k},2}\beta^{r\dag}_{-{\bf k},1})+
\\
 &+&\epsilon^{r}\sin^{2}\theta \;|V_{{\bf k}}||U_{{\bf
k}}|(\alpha^{r\dag}_{{\bf k},1}\beta^{r\dag}_{-{\bf k},1}-
\alpha^{r\dag}_{{\bf k},2}\beta^{r\dag}_{-{\bf k},2})
+\sin^{2}\theta \; |V_{{\bf k}}|^{2}\alpha^{r\dag}_{{\bf
k},1}\beta^{r\dag}_{-{\bf k},2} \alpha^{r\dag}_{{\bf
k},2}\beta^{r\dag}_{-{\bf k},1} \Big]|0\rangle_{1,2}\,.
 \eea
 Eq.(\ref{0emu}) exhibits the condensate
structure of the flavor vacuum $|0\rangle_{e,\mu}$. The important
point is that $_{1,2}\langle 0 |0 (t)\rangle_{e,\mu} \rar 0$, for any $t$,
in the infinite volume limit \cite{BV95}. Thus, in such a limit the
Hilbert spaces ${\cal H}_{1,2}$ and ${\cal H}_{e,\mu}$ turn out to
be unitarily inequivalent spaces. We remark that  $|0\rangle_{e,\mu}$ is the physical vacuum
as we will see below.

In the QM formalism we ''cannot'' have two unitarily inequivalent
Hilbert spaces for the simple reason that the von Neumann theorem
forbids the existence of unitarily inequivalent representations of
the canonical (anti-)commutation rules whenever the system has a
finite number of degrees of freedom, such as those in QM. It is
quite obvious that this cannot be the case for neutrinos. Since they
are quantum fields, by definition they are described by infinitely
many degrees of freedom and thus von Neumann theorem does not hold.
This point introduces a crucial difference between the QFT formalism
and the QM approach.

The flavor annihilators, relative
to the fields $\nu_{e}(x)$ and $\nu_{\mu}(x)$ at each time, are given by (we
use $(\sigma,i)=(e,1) , (\mu,2)$):
\begin{eqnarray}\label{flavannich}
\alpha _{{\bf k},\sigma}^{r}(t) &\equiv &G^{-1}_{\bf \te}(t)\;\alpha
_{{\bf k},i}^{r}(t)\;G_{\bf \te}(t)\,,  \nonumber
\\
\beta _{{\bf k},\sigma}^{r}(t) &\equiv &G^{-1}_{\bf \te}(t)\;\beta
_{{\bf
k},i}^{r}(t)\;G_{\bf \te}(t)\,.
\end{eqnarray}

The flavor fields can be expanded in the same bases as $\nu_{i}$:
\begin{eqnarray}
\nu _{\sigma}({\bf x},t) &=&\frac{1}{\sqrt{V}}{\sum_{{\bf k},r} }
e^{i{\bf k.x}}\left[ u_{{\bf k},i}^{r} \alpha _{{\bf
k},\sigma}^{r}(t) + v_{-{\bf k},i}^{r} \beta _{-{\bf k},\sigma}^{r\dagger
}(t)\right]\,.
\end{eqnarray}

The flavor annihilation operators
in the reference frame
such that ${\bf k}=(0,0,|{\bf k}|)$ are:
\bea\label{annihilator} \non
\alpha^{r}_{{\bf
k},e}(t)&=&\cos\theta\;\alpha^{r}_{{\bf
k},1}(t)\;+\;\sin\theta\;\left( |U_{{\bf k}}|\; \alpha^{r}_{{\bf
k},2}(t)\;+\;\epsilon^{r}\; |V_{{\bf k}}|\; \beta^{r\dag}_{-{\bf
k},2}(t)\right)
\\ \non
\alpha^{r}_{{\bf k},\mu}(t)&=&\cos\theta\;\alpha^{r}_{{\bf
k},2}(t)\;-\;\sin\theta\;\left( |U_{{\bf k}}|\; \alpha^{r}_{{\bf
k},1}(t)\;-\;\epsilon^{r}\; |V_{{\bf k}}|\; \beta^{r\dag}_{-{\bf
k},1}(t)\right)
\\
\beta^{r}_{-{\bf k},e}(t)&=&\cos\theta\;\beta^{r}_{-{\bf
k},1}(t)\;+\;\sin\theta\;\left( |U_{{\bf k}}|\; \beta^{r}_{-{\bf
k},2}(t)\;-\;\epsilon^{r}\; |V_{{\bf k}}|\; \alpha^{r\dag}_{{\bf
k},2}(t)\right)
\\ \non
\beta^{r}_{-{\bf k},\mu}(t)&=&\cos\theta\;\beta^{r}_{-{\bf
k},2}(t)\;-\;\sin\theta\;\left( |U_{{\bf k}}|\; \beta^{r}_{-{\bf
k},1}(t)\;+\;\epsilon^{r}\; |V_{{\bf k}}|\; \alpha^{r\dag}_{{\bf
k},1}(t)\right),
\eea
where $\epsilon^{r}=(-1)^{r}$ and
\bea
 |U_{{\bf k}}|\equiv u^{r\dag}_{{\bf k},i}
u^{r}_{{\bf k},j} = v^{r\dag}_{-{\bf k},i} v^{r}_{-{\bf k},j} \,
,\qquad  i \neq j\,; \qquad \qquad |V_{{\bf k}}|\equiv
\varepsilon_{ij} \epsilon^{r}\; u^{r\dag}_{{\bf k},i} v^{r}_{-{\bf
k},j}, \quad {\rm no \,\, summation} \eea
with $\varepsilon_{ij}= 0,1,-1$ for $i=j,~ i<j,  ~i>j$,
respectively.
We have:
  \bea
\non |U_{{\bf
k}}|=\left(\frac{\omega_{k,1}+m_{1}}{2\omega_{k,1}}\right)^{\frac{1}{2}}
\left(\frac{\omega_{k,2}+m_{2}}{2\omega_{k,2}}\right)^{\frac{1}{2}}
\left(1+\frac{{\bf
k}^{2}}{(\omega_{k,1}+m_{1})(\omega_{k,2}+m_{2})}\right)
\\
\label{Vk}|V_{{\bf
k}}|=\left(\frac{\omega_{k,1}+m_{1}}{2\omega_{k,1}}\right)^{\frac{1}{2}}
\left(\frac{\omega_{k,2}+m_{2}}{2\omega_{k,2}}\right)^{\frac{1}{2}}
\left(\frac{k}{(\omega_{k,2}+m_{2})}-\frac{k}{(\omega_{k,1}+m_{1})}\right)
\eea
\bea
|U_{{\bf k}}|^{2}+|V_{{\bf k}}|^{2}=1\,.
\eea

The  number of  condensate neutrinos for each $\bf k$ is given
by
\bea \label{density} _{e,\mu}\langle 0| \al_{{\bf k},i}^{r \dag}
\al^r_{{\bf k},i} |0\rangle_{e,\mu}\,= \;_{e,\mu}\langle 0|
\bt_{{\bf k},i}^{r \dag} \bt^r_{{\bf k},i} |0\rangle_{e,\mu}\,=\,
\sin^{2}\te\; |V_{{\bf k}}|^{2} \;,
 \eea
with $i=1,2\,$. Equivalently, $ _{1,2}\langle 0| \al_{{\bf
k},\sigma}^{r \dag} \al^r_{{\bf k},\sigma} |0\rangle_{1,2}\,=
\;_{1,2}\langle 0| \bt_{{\bf k},\sigma}^{r \dag} \bt^r_{{\bf
k},\sigma} |0\rangle_{1,2}\,=\, \sin^{2}\te\; |V_{{\bf k}}|^{2}
\;,$ with $\sigma = e,\mu\,$.

The Bogoliubov coefficient $|V_{{\bf k}}|^{2}$ appearing in
Eq.(\ref{density}) can be written as a function of the dimensionless
momentum $p=\frac{|{\bf k}|}{\sqrt{m_1 m_2}}$ and dimensionless
parameter $a= \frac{(m_{2}-m_{1})^2}{m_1 m_2}$, as follows,
\bea |V(p,a)|^2 & =& \frac{1}{2}\lf(1-\frac{p^2+1} {\sqrt{(p^2 +
1)^2 + a p^2}}\ri) ~.\label{Vpa}
 \eea

From Fig.1 we see that the effect is maximal when $p=1$, i.e. for
$|{\bf k}|^2 = m_{1} m_{2}$, the natural scale of the neutrino
mixing.  $|V_{{\bf k}}|^2$ goes to zero for large momenta, i.e. for
$|{\bf k}|^2\gg m_{1} m_{2}$,  as $|V_{{\bf k}}|^2 \approx
\frac{(\Delta m)^2}{4 k^2}$.   It acts as a ``form factor'' in the
$\bf k$ space controlling the neutrino vacuum condensate. We thus
find that the Pontecorvo's formalism is nothing but the relativistic
limit of the QFT formalism: $|V_{{\bf k}}|^2 \rar 0$ and $|U_{{\bf
k}}|^2 \rar 1$ for  $|{\bf k}|^2 \gg m_{1} m_{2}$. In the
Pontecorvo's formalism the non-perturbative contributions from the
vacuum condensate are thus missing. This is the meaning of the
approximation made in the QM treatment of the mixing. Missing these
condensate contributions is of course of no relevance for the
experimental observation of neutrino oscillations at today
instrumentation resolution. These contributions might play, however,
a relevant role in the study of the cosmological background. The
fact that $|V_{{\bf k}}|^2$ contributes maximally for low energies
suggests indeed to us that the contribution of the mixing phenomenon
may be taken as a good candidate in the study of dark energy.

\begin{figure}
\centering \resizebox{10cm}{!}{\includegraphics{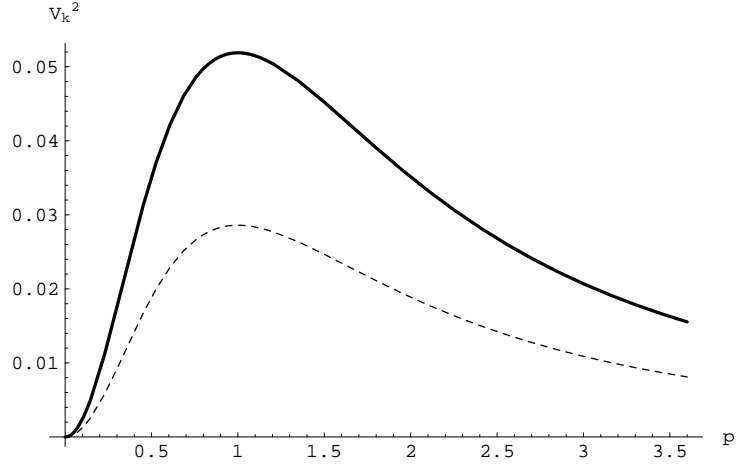}}
\caption{The fermion condensation density $|V(p,a)|^2$ as a
function of $p$ for $a=0.98$ (solid line) and $a=0.5$ (dashed
line).} \label{Fig: 1}
\end{figure}

We point out that, since $|0\rangle_{1,2}$ and $|0\rangle_{e,\mu}$
are unitary inequivalent states in the infinite volume limit, two
different normal orderings must be defined, respectively with
respect to the vacuum $|0\rangle_{1,2}$ for fields with definite
masses, denoted as usual  by $:...:$,  and with respect to the
vacuum for fields with definite flavor $|0\rangle_{e,\mu}$, denoted
by $::...::$~. The Hamiltonian normal ordered with respect to the
vacuum $|0\rangle_{1,2}\,$ is given by
\bea \label{Hnorm}:H: \,=\, H - _{1,2} \langle 0|H|0\rangle_{1,2} = H +
\, 2\int d^{3}{\bf k} \, (\omega_{k,2} + \omega_{k,1}) =
\sum_{i}\sum_{r}\int d^{3}{\bf k}\,\omega_{k,i}[\alpha_{{\bf
k},i}^{r \dag}\alpha_{{\bf k},i}^{r}+ \beta_{{\bf k},i}^{r
\dag}\beta_{{\bf k},i}^{r}] ~,
\eea
and the Hamiltonian normal ordered with respect
to the vacuum $|0\rangle_{e,\mu}$ is
\bea\label{Hflav} \nof H \nof\, \equiv \,  H \, -\, {}_{e,\mu}\lan
0(t)| H | 0(t) \ran_{e,\mu}\, = H \, +\, 2\int d^{3}{\bf k} \,
(\omega_{k,2} + \omega_{k,1}) \,(1 - 2\,|V_{\bf k}|^{2}
\sin^{2}\theta)\,. \eea
 Note that the difference of
energy between $|0\rangle_{e,\mu}$ and $|0\rangle_{1,2}$ represents
the energy of the condensed neutrinos given in Eq.(\ref{density})
\bea\label{energyCond} {}_{e,\mu}\lan 0(t)| :H: | 0(t)
\ran_{e,\mu}\,=\,
{}_{e,\mu}\lan 0(t)| H | 0(t) \ran_{e,\mu}\,-\, {}_{1,2}\lan 0| H | 0 \ran_{1,2}\,=\,
4\,\sin^{2}\theta\,\int d^{3}{\bf k} \, (\omega_{k,2} +
\omega_{k,1}) \,|V_{\bf k}|^{2} ~ , \eea
and gives the ``energy gap'' protecting the flavored neutrinos from
turning into the mixing component neutrinos  $\nu_1$ and $\nu_2$. In
the following we show that the energy of the condensed neutrinos can
have cosmological implications, indeed it can contribute to the dark
energy of the universe.

Before considering cosmological aspects of neutrino mixing, in order
to better understand the meaning of Eq.(\ref{energyCond}), let us
introduce the operator $A(t)$ that satisfies Eqs.(\ref{Aee}) -
(\ref{multi-A}) given in Appendix A. By defining the operator \bea
\label{H'} H'(t) \,=\, :H: - A(t)\,, \eea
 we have
\bea\label{H'(t)e}
\langle\nu_{{\bf k},e}^{r}(t)|\,
H'(t)\,|\nu_{{\bf k},e}^{r}(t)\rangle & = &
\omega_{k,1} \cos^{2}\theta\, + \omega_{k,2}\sin^{2}\theta\,,
\\\label{H'(t)mu}
\langle\nu_{{\bf k},\mu}^{r}(t)|\,
H'(t)\,|\nu_{{\bf k},\mu}^{r}(t)\rangle &=& \
\omega_{k,2} \cos^{2}\theta\, + \omega_{k,1}\sin^{2}\theta\,,
\\\label{H'(t)e-mu}
\langle\nu_{{\bf k},e}^{r}(t)|\,
H'(t)\,|\nu_{{\bf k},\mu}^{r}(t)\rangle &=&
(\omega_{k,2} - \omega_{k,1})\sin \theta \cos \theta\,,
\eea
\bea \label{H'(t)multi1}\non
&&\langle\nu_{{\bf k},\mu \bar{e} e}^{r}(t)|\,
H'(t)\,|\nu_{{\bf k},e}^{r}(t)\rangle \, = \,
\langle\nu_{{\bf k},\mu \bar{e} e}^{r}(t)|\,
H'(t)\,|\nu_{{\bf k},\mu}^{r}(t)\rangle\,= \,\langle\nu_{{\bf k},e
\bar{\mu} \mu}^{r}(t)|\,
H'(t)\,|\nu_{{\bf k},e}^{r}(t)\rangle \, = \,
\langle\nu_{{\bf k},e \bar{\mu} \mu}^{r}(t)|\,
H'(t)\,|\nu_{{\bf k},\mu}^{r}(t)\rangle = 0\,.
\\
\eea
Eqs.(\ref{H'(t)e})-(\ref{H'(t)e-mu}) coincide with the ones obtained
in QM by using the Pontecorvo states. Moreover the uncertainties in
the energy $H'(t)$ of the multi-particle states (\ref{multi1}),
(\ref{multi2}) are zero such as in QM, are zero the uncertainties in
the energy $H$ of the multi-particle states. $H'(t)$ is explicitly
given by
\bea \non\label{H'(t)} H'(t) & = & \sum_{r} \int d^{3}{\bf
k}\,\Big[\omega_{e e}
 \lf(\alpha_{{\bf k},e}^{r \dag}(t)\alpha_{{\bf k},e}^{r}(t)\, +
 \, \beta_{-{\bf k},e}^{r \dag}(t)\beta_{-{\bf k},e}^{r}(t) \ri)
\,+\,\omega_{\mu \mu} \lf(\alpha_{{\bf k},\mu}^{r \dag}(t)
 \alpha_{{\bf k},\mu}^{r}(t) + \beta_{-{\bf k},\mu}^{r \dag}(t)
 \beta_{-{\bf k},\mu}^{r}(t) \ri)
 \\
& + & \omega_{\mu e}
\lf(\alpha_{{\bf k},e}^{r \dag}(t)\alpha_{{\bf k},\mu}^{r}(t)\, +
\, \alpha_{{\bf k},\mu}^{r \dag}(t) \alpha_{{\bf k},e}^{r}(t)\, +
 \, \beta_{-{\bf k},e}^{r \dag}(t)\beta_{-{\bf k},\mu}^{r }(t)\, +
\, \beta_{-{\bf k},\mu}^{r \dag}(t) \beta_{{\bf k},e}^{r
}(t)\ri)\Big]\,, \eea
where $\omega_{e e} \equiv \omega_{k,1}\, \cos^{2}\theta \,+
\omega_{k,2}\, \sin^{2}\theta $, $\omega_{\mu \mu} \equiv
\omega_{k,1}\, \sin^{2}\theta \,+ \omega_{k,2}\, \cos^{2}\theta $, $
\omega_{\mu e} \equiv (\omega_{k,2} - \omega_{k,1})\,\sin\theta
\cos\theta $. From Eq.(\ref{H'(t)}) we have at any time $t$
\bea {}_{e,\mu}\lan 0(t)| H'(t) | 0(t) \ran_{e,\mu}\,=\, 0\,. \eea
Thus we obtain
\bea {}_{e,\mu}\lan 0(t)| :H: | 0(t) \ran_{e,\mu}\,
\equiv \, {}_{e,\mu}\lan 0(t)| A(t) | 0(t) \ran_{e,\mu}\,. \eea
That is, the operator $A(t)$ is the part of the Hamiltonian $:H:$
that give rise to the neutrino mixing condensate.
 The operator (\ref{H'(t)}) can be also written as
\bea
\label{H'}
H'(t) \,=\, \nof H \nof - B(t)\,,
\eea
where $B(t)$ satisfies Eqs.(\ref{Bee}) - (\ref{multi-B}) presented in Appendix A.

We remark that one might also consider the ``effective'' Hamiltonian
approach by incorporating into the QM treatment the condensate
contributions computed by using the operator $A(t)$ (or $B(t)$).

\section{Neutrino mixing and dark energy}

In this Section we show that
 the energy density of the
neutrino vacuum condensate can represent an evolving component of
the dark energy. The non-zero value of $|V_{\bf k}|^2$ for long
wavelengths, namely its behavior at very high momenta, together with
the negligible breaking of the Lorentz invariance of the vacuum
condensate at the present time, can be responsible of the very tiny
value of the cosmological constant.

Let us calculate the contribution $\rho_{vac}^{mix}$ of the
neutrino mixing to the vacuum energy density in the Minkowski metric.
 The energy-momentum tensor density
 ${\cal T}_{\mu \nu}(x) $ for the fields $\nu_1$ and $\nu_2$ is
\bea\
 :{\cal T}_{\mu \nu}(x): = \frac{i}{2}:\left({\bar \Psi}_{m}(x)\gamma_{\mu}
\stackrel{\leftrightarrow}{\partial}_{\nu} \Psi_{m}(x)\right):
\eea
where  $\Psi_{m} = (\nu_1, \nu_2)^{T}$. The symbol $:...:$ denotes
the normal ordering with respect to $|0 \rangle_{1,2}$.

In the early universe epochs, when the Lorentz invariance of the
vacuum condensate is broken,
 $\rho_{vac}^{mix}$ presents also space-time dependent condensate
 contributions. Then $\rho_{vac}^{mix}$ is given by
the expectation value of the (0,0) component of ${\cal T}_{\mu \nu}(x)$
in the flavor vacuum $|0(t){\rangle}_{e,\mu}$:
 \bea\
\rho_{vac}^{mix} = \frac{1}{V}\; \eta_{00}\; {}_{e,\mu}\lan 0(t)
| :T^{00}(0):| 0(t)\ran_{e,\mu}~,
 \eea
where $ :T_{00}:\,\equiv\,:H:\,=\,\int d^{3}x :{\cal T}_{00}(x): 
\,. $
Note that $T_{00}$ is time independent. 
We obtain
\bea\label{cc0}
 \rho_{vac}^{mix} = \frac{ 2}{\pi} \sin^{2}\theta
\int_{0}^{K} dk \, k^{2}(\omega_{k,1}+\omega_{k,2}) |V_{\bf
k}|^{2} \,,
\eea
where the cut-off $K$ has been introduced.
Explicitly
\bea\non\label{cc}
\rho_{vac}^{mix} &=& \frac{1}{2 \pi} \sin^{2}\theta (m_{2}-m_{1})
\Big\{ K \Big(m_{2} \sqrt{K^{2} + m_{2}^{2}} -
m_{1} \sqrt{K^{2} + m_{1}^{2}}\Big)
\\
&-&  m_{2}^{3}\log\lf(\frac{K+\sqrt{K^{2}+m_{2}^{2}}}{m_{2}}\ri)+
m_{1}^{3}\log\lf(\frac{K+\sqrt{K^{2}+m_{1}^{2}}}{m_{1}}\ri)
\Big\}\,.
\eea

The contribution
 $ p_{vac}^{mix}$ of the neutrino mixing
to the vacuum pressure is given by the expectation value of
${\cal T}_{jj}$ (where no summation on the index $j$ is intended)
in  $| 0\ran_{e,\mu}$:
 \bea\
p_{vac}^{mix}= -\frac{1}{V}\; \eta_{jj} \; {}_{e,\mu}\lan 0(t)
| :T^{jj}:| 0(t)\ran_{e,\mu} ~,
 \eea
 where $T_{jj} = \int d^{3}x \, {\cal T}_{jj}(x)\,. $
Being
 \bea\label{Tjj}
 :T^{jj}:= \sum_{i} \sum_{r}\int d^{3}{\bf k}\, \frac{k^j
k^j}{\;\omega_{k,i}}\lf(\al_{{\bf k},i}^{r\dag} \al_{{\bf
k},i}^{r}+ \beta_{{\bf -k},i}^{r\dag}\beta_{{\bf -k},i}^{r}\ri),
\eea
 in the case of the
isotropy of the momenta: $T^{11} = T^{22} = T^{33}$,  we have
 \bea\label{cc02}
  p_{vac}^{mix} = \frac{2}{3\;\pi}
\sin^{2}\theta \int_{0}^{K} dk \,
k^{4}\lf[\frac{1}{\omega_{k,1}}+\frac{1}{\omega_{k,2}}\ri] |V_{\bf
k}|^{2}\,.
 \eea
Explicitly
\bea\non\label{cc2}
p_{vac}^{mix} &=&
\frac{1}{6 \pi} \sin^{2}\theta (m_{2}-m_{1})
\Big\{ K \Big(m_{2} \sqrt{K^{2} + m_{1}^{2}} -
m_{1} \sqrt{K^{2} + m_{2}^{2}}\Big)
\\\non
&+&
2 \lf[ \frac{m_{2}^{4}}{\sqrt{m_{1}^{2}-m_{2}^{2}}}\, \arctan \Big(\frac{\sqrt{m_{1}^{2}-m_{2}^{2}}}{m_{2}\sqrt{K^{2} + m_{1}^{2}}}K \Big)
- \frac{m_{1}^{4}}{\sqrt{m_{2}^{2}-m_{1}^{2}}}\arctan \Big(\frac{\sqrt{m_{2}^{2}-m_{1}^{2}}}{m_{1}\sqrt{K^{2} + m_{2}^{2}}}K\Big)\ri]
\\
&+& (2 m_{1}^{3} + m_{1} m_{2}^{2})\log\lf(\frac{K+\sqrt{K^{2} + m_{2}^{2}}}{m_{2}}\ri)
- (2 m_{2}^{3} + m_{1}^{2}m_{2})\log\lf(\frac{K + \sqrt{K^{2} + m_{1}^{2}}}{m_{1}}\ri)
\Big\}\,.
\eea

The state equation of the vacuum condensate is defined as  $w^{mix} = p_{vac}^{mix}/
\rho_{vac}^{mix}$. By plotting $w^{mix} $ as function of the momentum cut-off $K$ (Fig.2) we have that
$w^{mix} = 1/3$ when the cut-off is chosen to be $K \gg m_{1}, m_{2}$ and $w^{mix}$
goes to zero for $K \leq \sqrt{m_{1} m_{2}}$.

\begin{figure}
\centering \resizebox{12cm}{!}{\includegraphics{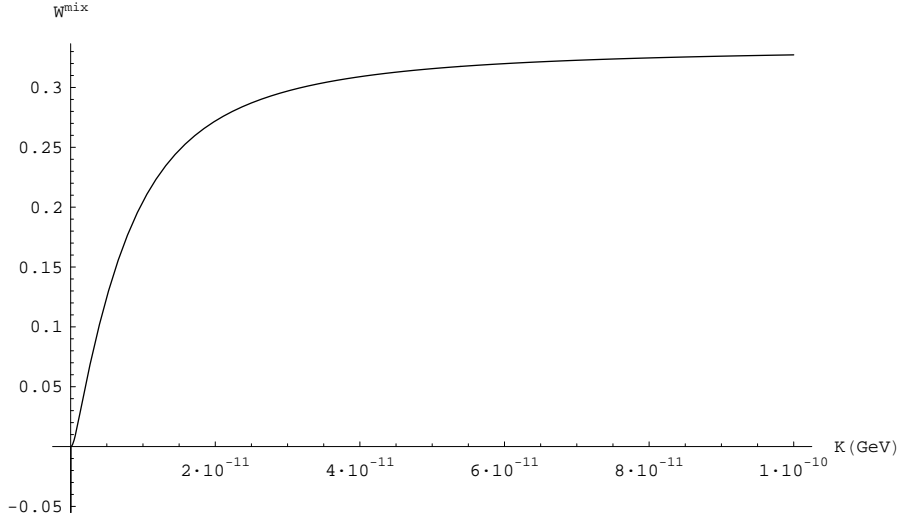}}
\hfill \caption{The adiabatic index $w^{mix}$ as a function of
cut-off K.} \label{Fig: 2}
\end{figure}

The neutrino vacuum condensate assume a different behavior at the present epoch.
Indeed, the breaking of the Lorentz invariance is now negligible and, at present time,
 $\rho_{vac}^{mix}$ comes from space-time independent condensate contributions.
Then the energy-momentum density tensor of the vacuum condensate
is given  approximatively by
\bea {}_{e,\mu}\lan 0 |:{\cal T}_{\mu\nu}:|
0\ran_{e,\mu} \approx \eta_{\mu\nu}\;\sum_{i}m_{i}\int
\frac{d^{3}x}{(2\pi)^3}\;{}_{e,\mu}\lan 0 |:\bar{\nu
}_{i}(x)\nu_{i}(x):| 0\ran_{e,\mu}\, =
\eta_{\mu\nu}\;\rho_{\Lambda}^{mix}.
 \eea

Since $\eta_{\mu\nu} = diag (1,-1,-1,-1)$ and, in a homogeneous and
isotropic universe, the energy-momentum tensor is ${\cal T}_{\mu\nu} = diag
(\rho\,,p\,,p\,,p\,)$,
the state equation is then $\rho_{\Lambda}^{mix} \approx -p_{\Lambda}^{mix}$,
that is, the neutrino vacuum condensate today has a  behavior similar to the
 cosmological constant {\cite{Capolupo:2006et}. Explicitly, we
have
\bea\label{cost} \rho_{\Lambda}^{mix}= \frac{2}{\pi} \sin^{2}\theta
\int_{0}^{K} dk \,
k^{2}\lf[\frac{m_{1}^{2}}{\omega_{k,1}}+\frac{m_{2}^{2}}
{\omega_{k,2}}\ri] |V_{\bf k}|^{2}. \eea

We observe that the value of the integral (\ref{cost}) is
conditioned by the appearance in the integrand of the $|V_{\bf
k}|^{2}$ factor. The integral, and thus $\rho_{\Lambda}^{mix}$,
would be zero for $|V_{\bf k}|^{2}= 0$ for any $|\bf k|$, as it is
in the usual quantum mechanical Pontecorvo formalism. In the present
QFT formalism $|V_{\bf k}|^{2}$ is not identically zero for any
$|\bf k|$; it goes to zero only for large momenta, getting its
maximum value for $|{\bf k}| = \sqrt{ m_{1} m_{2}} $ ($p =1$, cf.
Section II), thus maximally contributing in this (infrared) region:
we thus see that contributions to dark energy mostly come from long
wave-lengths, short wave-lengths at most producing local
dishomogeneities.
Proceeding in our calculation, we obtain
\bea\non\label{dark}
\rho_{\Lambda}^{mix} &=& \frac{1}{2 \pi} \sin^{2}\theta (m_{2}-m_{1})
\Big\{(m_{2} + m_{1}) K \Big(\sqrt{K^{2} + m_{2}^{2}}-
\sqrt{K^{2} + m_{1}^{2}}\Big)
\\\non
&+& 2
\Big[\frac{m_{1}^{4}}{\sqrt{m_{2}^{2}-m_{1}^{2}}}
\arctan \Big(\frac{\sqrt{m_{2}^{2}-m_{1}^{2}}}{m_{1}\sqrt{K^{2}+m_{2}^{2}}}K\Big)
- \frac{m_{2}^{4}}{\sqrt{m_{1}^{2}-m_{2}^{2}}} \arctan \Big(\frac{\sqrt{m_{1}^{2}-m_{2}^{2}}}{m_{2}\sqrt{K^{2}+m_{1}^{2}}}K\Big) \Big]
\\
&-& (m_{2}^{3} + 2 m_{1}^{3} + m_{1} m_{2}^{2})\log\lf(\frac{K+\sqrt{K^{2}+m_{2}^{2}}}{m_{2}}\ri)
+ (m_{1}^{3} + 2 m_{2}^{3} + m_{1}^{2}m_{2})\log\lf(\frac{K+\sqrt{K^{2}+m_{1}^{2}}}{m_{1}}\ri)
\Big\}.
\eea

\begin{figure}
\centering \resizebox{12cm}{!}{\includegraphics{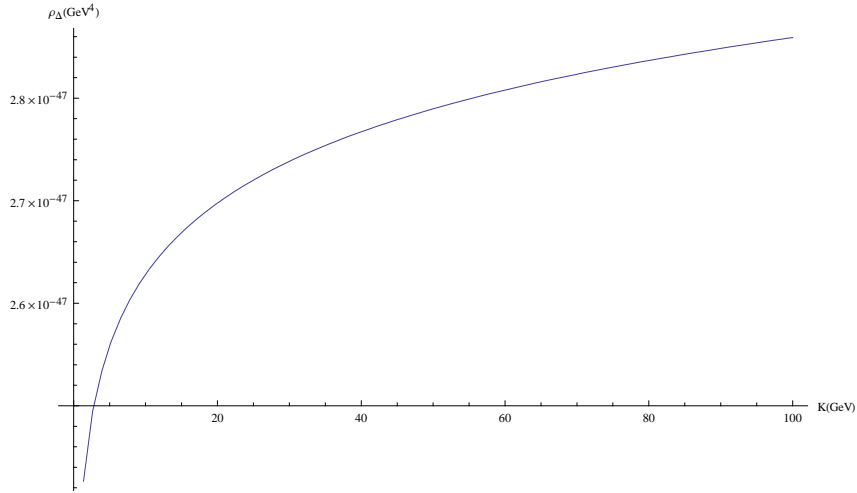}}
\hfill \caption{The neutrino mixing dark energy as a function of
cut-off K.} \label{Fig: 3}
\end{figure}

To better understand the meaning of Eq.(\ref{dark}), we report the behavior of
$\rho_{\Lambda}^{mix}$ for $K \gg m_{1},m_{2} $:
\bea\non\label{darklimit} \rho_{\Lambda}^{mix} & \approx &
\frac{1}{2 \pi} \sin^{2}\theta (m_{2}-m_{1})
\Big\{2
\Big[\frac{m_{1}^{4}}{\sqrt{m_{2}^{2}-m_{1}^{2}}}
\arctan
\Big(\frac{\sqrt{m_{2}^{2}-m_{1}^{2}}}{m_{1}}\Big) -
\frac{m_{2}^{4}}{\sqrt{m_{1}^{2}-m_{2}^{2}}} \arctan \Big(\frac{\sqrt{m_{1}^{2}-m_{2}^{2}}}{m_{2}}\Big)
\Big]
\\
 &-&  (m_{2}^{3} + 2 m_{1}^{3} + m_{1} m_{2}^{2})\log\lf(\frac{2K}{m_{2}}\ri)
+ (m_{1}^{3} + 2 m_{2}^{3} + m_{1}^{2}m_{2})\log\lf(\frac{2K}{m_{1}}\ri) \Big\}.
\eea
This shows that the integral diverges in $K$ as
$m_{i}^{4}\,\log\lf( 2K/m_{j} \ri)$, with $i,j = 1,2$. As shown in
Fig.3 the divergence in $K$ is smoothed by the factor $m_{i}^{4}$.
By using the electroweak scale cut-off: $K=100\, GeV,$ for neutrino masses of order of
$10^{-3} eV$   and $\Delta m_{12}^{2} \approx 7 \times 10^{-5} eV^{2}$
 we have $\rho_{\Lambda}^{mix} \approx 2.9 \times 10^{-47} GeV^{4}$,
 which is in agreement with the observed value of cosmological
constant. From Eq.
(\ref{darklimit}) one also sees that $\frac{d
\rho_{\Lambda}^{mix}(K)}{d K} \propto \frac{1}{K} \rightarrow 0$
for large $K$.

The result we have obtained is quite instructive also since it
tells us that the value of $|V_{\bf k}|^{2}$, for any $|\bf k|$,
contributing to the observed value of $\rho_{\Lambda}^{mix}$ is
the one related with such a mass scale (the dependence of $|V_{\bf
k}|^{2}$ on the masses is shown in Eq.(\ref{Vk}), see also Fig.1).
The computation of $\rho_{\Lambda}^{mix}$ turns out to be
sensible to small variations in the values of the neutrino masses
and of $\Delta m^{2}$,  these
last ones affecting the value of the multiplicative digits of
$10^{-47}GeV^{4}$.

Above we have derived the contribution  $\rho_{\Lambda}^{mix}$
arising from mixing of the two lighter neutrinos. If mixing involving
heaviest neutrinos are included, the obtained value for dark energy, for a value of the
cut-off of order of the electroweak scale, turns out
 about $4$  orders of magnitudes higher than the observed
value of dark energy.
In such a case, a value of the dark energy compatible
with its upper bound is obtained for a cut-off of the order of the natural scale of
the neutrino mixing. Such a small cut off on the momenta is imposed by
 the negligible Lorentz invariance breaking at the present epoch, as we will show
in a forthcoming paper \cite{Capolupo2008}.
 There we will present also the
explicit computation in curved space-time and we will show that
the mixing treatment here presented in the flat space-time  is a
good approximation in the present epoch of that in FRW space-time.

\section{Conclusions}

In this report we have presented the main features of neutrino mixing
 in the context of quantum field theory and
we have shown that neutrino mixing may contribute
to the value of the dark energy exactly
because of the non-perturbative  field theory effects.
In particular, we have shown that, at the present epoch, the vacuum condensate generated by neutrino mixing behaves as the cosmological constant.
Its observed value is obtained for a cut-off of the order of electroweak scale
when the two lighter neutrinos are considered
and for a cut-off of the order of the natural scale of neutrino mixing
in the case one includes mixing with heavier neutrinos.

\section*{Acknowledgements}

Support from INFN and MIUR is acknowledged.

\appendix

\section{Expectation values of the operators A, and B}

The operator $A(t)$ satisfies the relations:
\bea
\label{Aee}
\langle\nu_{{\bf k},e}^{r}(t)|\,A(t) \,|\nu_{{\bf k},e}^{r}(t)\rangle &=&
2 \omega_{k,1} |V_{\bf k}|^{2}\sin^{2}\theta\,;
\qquad
\langle\nu_{{\bf k},\mu}^{r}(t)|\,A(t)\,|\nu_{{\bf k},\mu}^{r}(t)\rangle\, =\,
2 \omega_{k,2} |V_{\bf k}|^{2}\sin^{2}\theta\,,
\\
\langle\nu_{{\bf k},e}^{r}(t)|\, A(t)\,|\nu_{{\bf k},\mu}^{r}(t)\rangle &=&
\langle\nu_{{\bf k},\mu}^{r}(t)|\, A(t)\,|\nu_{{\bf k},e}^{r}(t)\rangle =
\lf(\omega_{k,2}-\omega_{k,1}\ri)(|U_{\bf k}|-1)\sin \theta \cos \theta\,,
\eea
and similar relations hold for the anti-particle states
$|\overline{\nu}_{\sigma}(t)\rangle $,
moreover
\bea
\langle\nu_{{\bf k},e \bar{\mu} \mu}^{r}(t)|\,
A(t)\,|\nu_{{\bf k},e}^{r}(t)\rangle & = &
2\, \epsilon^r \,\omega_{k,1}\,\sin^{2}\theta\,|U_{\bf k}|\,|V_{\bf k}|\,;
\quad\,\,\,
 \langle\nu_{{\bf k},{\mu} \bar{e} e}^{r}(t)|\,
A(t)\,|\nu_{{\bf k},\mu}^{r}(t)\rangle \, = \,
-2\, \epsilon^r \,\omega_{k,2}\,\sin^{2}\theta\,|U_{\bf k}|\,|V_{\bf k}|,
\\ \label{multi-A}
\langle\nu_{{\bf k},e \bar{\mu} \mu}^{r}(t)|\,
A(t)\,|\nu_{{\bf k},\mu}^{r}(t)\rangle & = & \langle\nu_{{\bf k},{\mu} \bar{e} e}^{r}(t)|\,
A(t)\,|\nu_{{\bf k},e}^{r}(t)\rangle  \,=\,
\epsilon^r \,\lf(\omega_{k,2}+\omega_{k,1}\ri)\,|V_{\bf k}|\,\sin \theta \,\cos \theta\,,
\eea
where, at time $t$, the multi-particle flavor states are defined as:
\bea
\label{multi1} |\nu_{{\bf k},e \bar{e} \mu}^{r}(t)\rangle & \equiv &
 \alpha_{{\bf k},e}^{r \dag}(t)\,
\beta_{-{\bf k},e}^{r \dag}(t)\, \alpha_{{\bf k},\mu}^{r
\dag}(t)\,|0(t)\rangle_{e,\mu} \,,
\\ [2mm]
\label{multi2}|\nu_{{\bf k},\mu  \bar{\mu} e}^{r}(t)\rangle & \equiv &
\alpha_{{\bf k},\mu}^{r \dag}(t)\, \beta_{-{\bf k},\mu}^{r \dag}(t)\,
\alpha_{{\bf k},e}^{r \dag}(t)\,|0(t)\rangle_{e,\mu}\,.
\eea

The operator B in Eq.(\ref{H'}) has the expectation values given below:
\bea\label{Bee}
\langle\nu_{{\bf k},e}^{r}(t)|\,B(t) \,|\nu_{{\bf k},e}^{r}(t)\rangle &=&
-2 \omega_{k,2} |V_{\bf k}|^{2}\sin^{2}\theta\,;
\qquad
\langle\nu_{{\bf k},\mu}^{r}(t)|\,B(t)\,|\nu_{{\bf k},\mu}^{r}(t)\rangle \,=\,
-2 \omega_{k,1} |V_{\bf k}|^{2}\sin^{2}\theta\,,
\\
\langle\nu_{{\bf k},e}^{r}(t)|\, B(t)\,|\nu_{{\bf k},\mu}^{r}(t)\rangle &=&
\langle\nu_{{\bf k},\mu}^{r}(t)|\, B(t)\,|\nu_{{\bf k},e}^{r}(t)\rangle =
\lf(\omega_{k,2}-\omega_{k,1}\ri)(|U_{\bf k}|-1)\sin \theta \cos \theta\,,
\eea
and similar for $|\overline{\nu}_{\sigma}(t)\rangle $, moreover
\bea
\langle\nu_{{\bf k},e \bar{\mu} \mu}^{r}(t)|\,
B(t)\,|\nu_{{\bf k},e}^{r}(t)\rangle & = &
2\, \epsilon^r \,\omega_{k,1}\,\sin^{2}\theta\,|U_{\bf k}|\,|V_{\bf k}|\,;
\quad\,\,\,
\langle\nu_{{\bf k},\mu \bar{e} e}^{r}(t)|\,
B(t)\,|\nu_{{\bf k},\mu}^{r}(t)\rangle \, = \,
-2\, \epsilon^r \,\omega_{k,2}\,\sin^{2}\theta\,|U_{\bf k}|\,|V_{\bf k}|,
\\ \label{multi-B}
\langle\nu_{{\bf k},e \bar{\mu} \mu}^{r}(t)|\,
B(t)\,|\nu_{{\bf k},\mu}^{r}(t)\rangle & = &\langle\nu_{{\bf k},\mu \bar{e} e}^{r}(t)|\,
B(t)\,|\nu_{{\bf k},e}^{r}(t)\rangle\,=\,
\epsilon^r \,\lf(\omega_{k,2}+\omega_{k,1}\ri)\,|V_{\bf k}|\,\sin \theta \,\cos \theta\,.
\eea


\end{document}